\documentclass[twocolumn,showpacs,preprintnumbers,amsmath,amssymb]{revtex4-1}

\usepackage{graphicx}
\usepackage{dcolumn}
\usepackage{bm}

\def\bequ{\begin{equation}}
\def\eequ{\end{equation}}
\def\be{\begin{equation}}
\def\ee{\end{equation}}
\def\nn{\nonumber}
\def\del{\partial}

\begin{document}


\title{$n$-DBI gravity, maximal slicing and the Kerr geometry}

\author{Fl\'avio S. Coelho}
 \email{flavio@physics.org}
\author{Carlos Herdeiro}%
\author{Mengjie Wang}%
\affiliation{\vspace{2mm}Departamento de F\'\i sica da Universidade de Aveiro and I3N \\ 
Campus de Santiago, 3810-183 Aveiro, Portugal \vspace{1mm}}%

\date{October 2012}

\begin{abstract}
Recently \cite{Herdeiro:2011im}, we have established that solutions of Einstein's gravity admitting foliations with a certain geometric condition are also solutions of $n$-DBI gravity \cite{Herdeiro:2011km}. Here we observe that, in vacuum, the required geometric condition is fulfilled by the well known \textit{maximal slicing}, often used in numerical relativity. As a corollary, we establish that the Kerr geometry is a solution of $n$-DBI gravity in the foliation adapted to Boyer-Lindquist coordinates.
\end{abstract}

\maketitle

There are at present important observational motivations to explore relativistic gravity beyond General Relativity (GR). Firstly, there is a large body of Cosmological evidence for an accelerating Universe \cite{Weinberg:2012es}, which, within GR, requires an exotic form of energy, raising the question if such exotic energy could be traded by a modification in the laws of gravity. Secondly, the expected opening of the gravitational wave astronomy field in the very near future, with the planned science runs for the second generation of gravitational wave observatories - such as Advanced LIGO - will test GR and constrain alternative theories of gravity in new ways. Indeed, already the observation of the orbital period variation in binary pulsar systems, in particular those with a large mass ratio between the neutron stars, has set important constraints in scalar-tensor and TeVeS theories of gravity \cite{Freire:2012mg}. 

Most alternative theories of gravity explored for phenomenological purposes keep the central diffeomorphism invariance of GR; but models with preferred reference frames have also been considered, a notable example being the Einstein-aether theory  \cite{Jacobson:2000xp}.  In 2009, Ho\v rava proposed a non fully covariant theory of gravity, as a means to address the well known non-renormalizability property of GR \cite{Horava:2009uw}, which has drawn attention to gravity theories where the symmetry group is reduced to the so-called Foliation Preserving Diffeomorphisms (FPD). One such model that has been recently proposed is $n$-DBI gravity \cite{Herdeiro:2011km,Herdeiro:2011im,Coelho:2012xi}. This model, albeit not proposed as a fundamental (i.e. quantum) theory of gravity, has appealing properties, such as equations of motion that are higher order in spatial derivatives, but only second order in time derivatives, which would avoid loss of unitarity in the quantum regime. Moreover, it revealed interesting properties for phenomenological studies: it can explain early and late time inflation in a unified way \cite{Herdeiro:2011km}; it contains as solutions the standard spherically symmetric black hole geometries of GR, appropriately foliated \cite{Herdeiro:2011im}; it is not afflicted by the same pathologies as the initial Ho\v rava-Lifshitz proposal, namely, instabilities, strong coupling problems and absence of lapse dynamics \cite{Coelho:2012xi}. It therefore merits further study as a learning ground for phenomenology beyond GR. 

A way to learn about the properties of any model of gravity is to consider its exact solutions. One simple question concerning $n$-DBI gravity, raised in \cite{Herdeiro:2011im}, is if the Kerr solution of vacuum GR, appropriately foliated, is also a solution of $n$-DBI. This question becomes even more interesting if we notice 
that only slowly rotating black holes have been (recently) found in Ho\v rava-Lifschitz gravity \cite{Barausse:2012qh,Wang:2012nv}.
%
The purpose of this note is to answer affirmatively to such question by noting that the geometric condition found in \cite{Herdeiro:2011im} for solutions of GR to be solutions of $n$-DBI has a clear interpretation in the theory of space-time foliations.


$n$-DBI gravity (without matter) is described by the action \cite{Herdeiro:2011km}
\bequ
S=-\frac{3\lambda}{4\pi G_N^2}\int d^4x \sqrt{-g}\left[\sqrt{1+\frac{G_N}{6\lambda}\mathcal{R}}-q\right]\ \ ,
\eequ
where $\lambda,q$ are two constants, and 
\bequ
\mathcal{R}=\mbox{}^{\mbox{}^{(4)}\!}\! R-2 D_\mu(n^\mu D_\nu n^\nu) \ , \label{curlyr}
\eequ
 is the sum of the four dimensional Ricci scalar with a total derivative term closely related to the Gibbons-Hawking-York boundary term \cite{York:1972sj,Gibbons:1976ue}. The unit vector ${\bf n}$ is time-like and defines a preferred foliation of space-time; $D_\mu$ is the four dimensional covariant derivative.

A fundamental distinction between the solutions of this theory, as compared to GR or, say, $f(R)$ gravity theories, is that they are defined by their intrinsic geometry - encoded in the four dimensional curvature - \textit{and} the foliation chosen. The complexity of the field equations discourages an attempt at finding even the most generic spherically symmetric solution, which does not appear to imply staticity in this model (i.e. Birkhoff's theorem does not hold), by directly tackling these equations. But a considerable simplification occurs if we focus on the special case with constant $\mathcal{R}$, as we now describe.

First we observe that making a 3+1 ADM decomposition adapted to an Eulerian observer tangent to ${\bf n}$, we have
\be
\mathcal{R}=R+K_{ij}K^{ij}-K^2-2N^{-1}\Delta N \ , \label{defr}
\ee
where $N$ is the lapse, $K_{ij}$ the extrinsic curvature and $R,\Delta$ the 3-dimensional Ricci scalar and Laplacian.

Then, denoting 
\be
 C\equiv \sqrt{1+\frac{G_N}{6\lambda}\mathcal{R}} \ , \label{c}
\ee
the field equations reduce to
\begin{align}
\hspace{-.3cm}
R-N^{-1}\Delta N+\frac{6\lambda}{G_N}(1-qC)=0 \ ,\label{NvariationR}
\end{align}
\be
\nabla^j\left({K_{ij}-h_{ij}K }\right)=0\ ,
\ee
\begin{align}
\hspace{-.3cm}
\pounds_{\bf n}K_{ij}= 
\frac{\nabla_i\nabla_j N}{N}-R_{ij}-KK_{ij}+2K_{il}K_j^{\ l}+h_{ij}G_N\Lambda_C \ , \label{gammavariationR}
\end{align}
where $\pounds_{\bf n} =\frac{1}{N}(\partial_t-\pounds_N)$ is the Lie derivative along $n$ and $\pounds_N$ is the Lie derivative along the shift.  The Hamiltonian constraint, momentum constraints and the dynamical equations are those of Einstein gravity with a cosmological constant
\be
\Lambda_{C}=\frac{3\lambda}{G_N^2}(2qC-1-C^2) \ . \label{lambdac}
\ee
It follows, as observed in \cite{Herdeiro:2011im}, that any solution of Einstein's gravity plus a cosmological constant admitting a foliation with constant $\mathcal{R}$ is a solution of $n$-DBI gravity. 

This slicing property can be rephrased in terms of (\ref{NvariationR}). The foliation is such that 
\be
R-N^{-1}\Delta N=\frac{6\lambda}{G_N}(qC-1)={\rm constant} \ . 
\label{slicing}
\ee
Let us also note that using the evolution equation for $K_{ij}$ and the Hamiltonian constraint one arrives at the following equation for the evolution of the trace of the extrinsic curvature:
\be
\frac{1}{N}(\partial_t-\pounds_N)K=-K^2+\frac{\Delta N}{N}-R+3G_N\Lambda_C \ .
\label{evok}
\ee

Educated observation of equations (\ref{slicing}) and (\ref{evok}) unveils the following fact: choosing the \textit{maximal slicing condition} for the foliation, defined as $K=0=\partial_t K$ \cite{Alcubierre:2008}, which clearly implies $\pounds_{\bf n}K=0$, requires, from eq. (\ref{evok}),
\be
R-N^{-1}\Delta N=3G_N\Lambda_C \ .
\ee
Maximal slicing is interpreted as the requirement that the volume element associated to Eulerian observers remains constant. This slicing is typically considered (mostly in the field of numerical relativity) in asymptotically flat space-times  - in a cosmological space-time volume elements naturally change with time. Thus we will focus on the case with $\Lambda_C=0$. Then,  (\ref{lambdac}) and (\ref{slicing}), determine $q$ and $C$ to be $q=C=1$, or, equivalently, $\mathcal{R}=0$.  For vacuum solutions of the Einstein equations this means that $D_\mu(n^\mu D_\nu n^\nu)=0 $ which is indeed verified as a consequence of $K=D_\nu n^\nu=0$. We thus arrive at the main message of this note:

\textit{Any vacuum solution of Einstein's equations in a foliation obeying the maximal slicing condition is a solution of $n$-DBI gravity.}
\bigskip

There is a simple and nice way to double check this result \cite{hirano}. In \cite{Coelho:2012xi} it was observed that the $n$-DBI action could be linearised, by introducing an auxiliary field $e$; then the action takes an Einstein-Hilbert form (albeit in a Jordan frame)
\begin{equation}
S^{e}=-\frac{1}{16\pi G_N}\int d^4x\sqrt{-g}e\left[\mathcal{R}-2G_N\Lambda_C(e)\right]\ ,
\label{linearisedaction}
\end{equation}
where $\Lambda_C(e)$ is given by (\ref{lambdac}) with $C\rightarrow 1/e$ in the right hand side. This  reduces to GR coupled to a cosmological constant if the auxiliary field is constant. Since the auxiliary field equation of motion yields (\ref{c}) with $C\rightarrow 1/e$, the constancy of $e$ is equivalent to the constancy of $\mathcal{R}$. In vacuum $\mbox{}^{\mbox{}^{(4)}\!}\! R=0$ and thus, from \eqref{curlyr}, $D_\mu(n^\mu D_\nu n^\nu)$ must be constant, for which maximal slicing is a sufficient condition.

The above result gives a concrete handle to obtain explicit solutions to $n$-DBI gravity, by invoking the literature on maximal slicing. Let us first illustrate this by reconsidering the spherically symmetric solutions.

In \cite{Herdeiro:2011im} a family of spherically symmetric, time independent solutions to $n$-DBI was obtained, of which the relevant sub-set we wish to consider reads:
\begin{align}
& ds^2=  -\left(1-\frac{2G_N M_1}{r}+\frac{C_3}{r^4}\right)dT^2+\nn \\
& \left(\frac{dr}{\sqrt{1-\frac{2G_NM_1}{r}+\frac{C_3}{r^4}}}+\sqrt{\frac{2G_NM_2}{r}+\frac{C_3}{r^4}}dT\right)^2+r^2d\Omega_2 \, . \label{dRNdS}
\end{align}
As a four dimensional geometry, this is simply the Schwarzschild manifold, with mass $M=M_1+M_2$, and can be transformed into Schwarzschild coordinates $(t,r,\theta,\phi)$ by the \textit{non-FPD}: 
\be
dt=dT-\frac{1}{1-\frac{2G_NM}{r}}\sqrt{\frac{\frac{2G_NM_2}{r}+\frac{C_3}{r^4}}{1-\frac{2G_NM_1}{r}+\frac{C_3}{r^4}}}dr \ . \label{ct}
\ee
But taking as equivalent solutions only those related by FPDs, these \textit{Schwarzschild geometries} are, generically, inequivalent. Moreover, scalar quantities constructed from the extrinsic curvature are now curvature invariants, since they are preserved by FPDs. The `geometric invariant' $K$ reads:
\be
K=-\frac{3 G_NM_2 }{\sqrt{C_3+2 G_NM_2 r^3}}\ ,
\ee
and therefore taking $M_2=0$ we obtain a family of Schwarzschild geometries, parameterised by $C_3$, \textit{all maximally sliced}. These are described in the literature \cite{Alcubierre:2008}, where the constant $C_3$ parameterising the different maximal slicings is the \textit{Estabrook-Wahlquist time} \cite{Estabrook:1973ue}.

We now turn our attention to the Kerr geometry. Start by noting that in the standard Boyer-Lindquist coordinates $(t,r,\theta,\phi)$, the Kerr geometry is maximally sliced. To realize this it is enough to notice that, in these coordinates, the ADM form of the Kerr geometry has a lapse, shift and 3-metric of the form:
\be
N=N(r,\theta) \ , \ \ \ \ \  N^i\partial_i=N^\phi(r,\theta)\partial_\phi \ , \ \ \ \ \ h_{ij}=h_{ij}(r,\theta) \ .
\ee
The natural foliation introduced by these coordinates, orthogonal to $n^\mu=(1/N,-N^i/N)$, has therefore extrinsic curvature with trace
\be
K=D_\mu n^\mu=-\frac{1}{N\sqrt{h}}\partial_i\left(\sqrt{h}N^i\right)=0 \ .
\ee
Thus, this is a maximal slicing and therefore yields a solution of $n$-DBI gravity (since it is a solution of GR).

In principle, by applying a non-FPD to the Kerr geometry in Boyer-Lindquist coordinates, and imposing the preservation of maximal slicing yields a \textit{family of Kerr geometries}, which solve the equations of $n$-DBI gravity. Concretely, we should redefine the time coordinate $t\longrightarrow T=t-H(r,\theta)$, where $H(r,\theta)$ is known as the \textit{height function}, build the normal unit vector $n_\mu=-N D_\mu T$, with $N$ determined by the normalization condition of ${\bf n}$, and finally 
add the condition of vanishing $K$, $\del_\mu(\sqrt{-g}n^\mu)=0$.
One then finds an explicit PDE whose solutions yield a set of Kerr geometries. It would be interesting - and indeed has been attempted, mostly by the numerical relativity community - to find explicit solutions to this PDE. But no explicit analytic solutions are known and finding them seems by no means a straightforward task.



The analysis already made in \cite{Herdeiro:2011im}, complemented by the observation made herein, shows that any vacuum solution of GR in a maximal slicing is a solution to $n$-DBI. The potential relevance of this observation depends, of course, on the physical interpretation of the solutions in $n$-DBI gravity. In particular: is this Kerr geometry a black hole in some meaningful way? Indeed the concept of a black hole as a trapped region, causally disconnected from some appropriate `exterior' requires an understanding of all propagating degrees of freedom. In models with a breakdown of Lorentz invariance, this issue requires detailed analysis, since different modes of the gravitational field may obey different dispersion relations.  In $n$-DBI gravity, however, the extra scalar graviton mode, which exists besides the two usual tensorial graviton modes of GR, does not seem to propagate \cite{Coelho:2012xi}, and hence, we may tentatively conclude that this Kerr \textit{geometry} indeed describes a Kerr \textit{black hole} in $n$-DBI gravity, as in GR.

Let us briefly comment on foliations with constant but non-vanishing $K$. If we take $q$ to have the value $q=1+G_N^2\Lambda_C/(6\lambda)$, required by the Einstein gravity limit \cite{Herdeiro:2011im}, we find, from (\ref{lambdac}), that $C=1$ or $C=1+G_N^2\Lambda_C/(3\lambda)$. Then, using  (\ref{slicing}), (\ref{evok}) and the constancy of $K$ we get $K^2=2G_N \Lambda_C$ or $K^2=-G_N^3 \Lambda_C^2/(3\lambda)$. Phenomenology of $n$-DBI gravity requires a positive $\lambda$ \cite{Herdeiro:2011km}; thus we conclude that \textit{an Einstein space foliated with constant $K$ yields a solution of $n$-DBI gravity (with the chosen $q$) if $K^2=2G_N \Lambda_C$}. It can be verified that the foliation of Kerr-(A)dS naturally induced by Boyer-Lindquist coordinates does not obey this constraint. 

In closing, let us observe that Schwarzschild-dS in a McVittie slicing has $K^2=3G_N \Lambda_C$ \cite{Zilhao:2012bb}; this provides a solution of $n$-DBI gravity with constant $\mathcal{R}$, but for $q=1/C$, rather than the aforementioned requirement.

\noindent{\bf{\em Acknowledgements.}}
 We would like to thank S. Hirano for discussions. F.C. and M.W.  are funded by FCT through the grants SFRH/BD/60272/2009 and SFRH/BD/51648/2011. The work in this paper is also supported by the grants PTDC/FIS/116625/2010 and  NRHEP--295189-FP7-PEOPLE-2011-IRSES.

\bibliographystyle{h-physrev4}
\bibliography{nDBI}


\end{document}